\documentclass[aps,prb,superscriptaddress,twocolumn,showpacs,preprintnumbers,amsmath,amssymb]{revtex4}
\usepackage{color}
\usepackage{array}
\usepackage{amsmath}
\usepackage{amssymb}
\usepackage{epsfig}
\usepackage{graphicx}
\usepackage{bbm}

\makeatletter

\newcommand{\Rmnum}[1]{\expandafter\@slowromancap\romannumeral #1@}
\makeatother

\begin{document}
\title{Possible interaction driven topological phases in (111) bilayers of LaNiO$_3$}

\author{Kai-Yu Yang}
\affiliation{Department of Physics, Boston College, Chestnut Hill,
Massachusetts 02467, USA}

\author{Wenguang Zhu}
\affiliation{Materials Science and Technology Division, Oak Ridge National Laboratory, Oak Ridge, Tennessee 37831, USA}
\affiliation{Department of Physics and Astronomy, University of Tennessee, Knoxville, Tennessee 37996, USA}
\author{Di Xiao}, \author{Satoshi Okamoto}
\affiliation{Materials Science and Technology Division, Oak Ridge National Laboratory, Oak Ridge, Tennessee 37831, USA}
\author{Ziqiang Wang}\author{Ying Ran}
\affiliation{Department of Physics, Boston College, Chestnut Hill,
Massachusetts 02467, USA}
\date{\today}

\begin{abstract}
We use the variational mean-field approach to systematically study the phase diagram of a bilayer heterostructure  of the correlated transition metal oxide LaNiO$_3$, grown along the (111) direction. The Ni$^{3+}$ ions with $d^7$ (or $e_g^1$) configuration form a buckled honeycomb lattice. We show that as a function of the strength of the on-site interactions, various topological phases emerge. In the presence of a reasonable size of the Hund's coupling, as the correlation is tuned from intermediate to strong, the following sequence of phases is found: (1) a Dirac half-semimetal phase, (2) a quantum anomalous Hall insulator (QAHI) phase with Chern number one, and (3) a ferromagnetic nematic phase breaking the lattice point group symmetry. The spin-orbit couplings and magnetism are both dynamically generated in the QAHI phase.

\end{abstract}
\maketitle

\emph{Introduction---} Artificial transition metal oxide heterostructures (TMOH) are becoming available owing to the recent development \cite{Izumi200153,Ohtomo2002,Ohtomo2004} in the fields of oxide superlattices and oxide electronics. In particular, layered structures of TMOH can now be prepared with atomic precision, thus offering a high degree of control over important material properties, such as lattice constant, carrier concentration, spin-orbit coupling, and correlation strength. The previous efforts on TMOH has been mainly focused on the (001) interface, where a rich variety of behavior emerges, such as superconductivity and magnetism(for a review, see Ref.\cite{Mannhart26032010}). In addition, recent theoretical investigation\cite{xiao-2011} pointed out that the bilayer TMOH grown along the (111) direction are promising materials realizing various topological phases.  

\begin{figure}
 \includegraphics[width=0.45\textwidth]{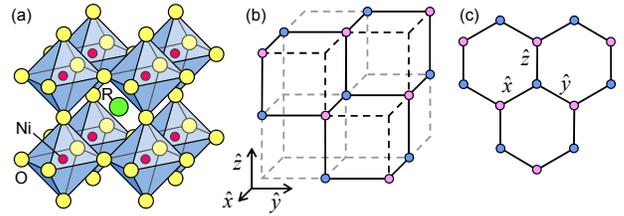}
\caption{(Color online) The structure of the (111) bilayer of RNiO3. (a) The original perovskite structure.
(b) The (111) bilayer of RNiO3. Here, only Ni sites are shown. (c) Buckled honeycomb lattice formed in the (111) bilayer. }
\label{fig:bilayer}
\end{figure}

The transition metal ions form a bulked honeycomb lattice in (111) bilayer structures (Fig.\ref{fig:bilayer}). Haldane first proposed that electrons hopping on a honeycomb lattice could realize the quantum Hall effect (QHE)
in the absence of Landau levels\cite{PhysRevLett.61.2015}, pointing out the possibility of nontrivial topology in simple band insulators. Such an insulator phase has been termed as quantum anomalous Hall insulator(QAHI). Its time-reversal symmetric generalization, topological insulators, have attracted a lot of interest both theoretically and experimentally (for reviews see Ref.\cite{RevModPhys.82.3045,2010arXiv1008.2026Q,Moore_Review_2010}). These insulators all feature a band gap driven by the spin-orbit coupling. In order to realize these physics experimentally, semimetallic materials are required. Honeycomb lattice is well-known to support semimetallic band structures, for instance in graphene. Therefore the TMOH along the (111) direction is particularly promising in searching for topological phases.

The perovskite nickelates, RNiO$_3$ where R is a rare-earch atom, have demonstrated rich physics including metal-insulator transitions. One very interesting feature of these systems is a rather complex pattern of charge and spin orders(for a recent summary of experimental progresses, see Ref.\cite{2011arXiv1107.0724L}). When R=La, the bulk compound remains metallic at all temperature. At low temperature it has a magnetic ordering pattern with an ``up-up-down-down'' spin configuration, coexisting with  a ``rock salt'' type charge order. The charge order has been argued to be a by-product of the spin order based on symmetry considerations\cite{PhysRevLett.106.016405}. The magnetic ordering pattern can be visualized in the following way\cite{PhysRevLett.106.016405}. When the cubic perovskite LaNiO$_3$ is viewed from the (111) direction, the Ni atoms form layers of triangular lattice. Each layer of the Ni atoms are ferromagnetically ordered. When these layers are stacked along the (111) direction, the periodicity of the pattern is four: i.e., ``up-up-down-down''. Namely two adjacent layers are both spin up and the next two adjacent layers are spin down. After including the orthorhombic distortion of the 3D lattice, this magnetic pattern is consistent with experimental observations\cite{PhysRevLett.106.016405}. Recently LaNiO$_3$/LaAlO$_3$ superlattices along the (001) direction have been actively investigated experimentally and display unique quantum confinement effects\cite{PhysRevB.83.161102}. 

Motivated by this interesting material, together with the recent experimental progresses of the growth of the (111) perovskite heterostructures (e.g. Ref\cite{10.1063/1.3525578}), we study the possible quantum phases of LaNiO$_3$ bilayer TMOH grown on insulating substrates such as LaAlO$_3$,LaScO$_3$.

\emph{Model---} Because LaNiO$_3$ bulk material is metallic, a correlated itinerant electronic model would be a reasonable starting point. The bilayer forms a buckled honeycomb lattice (see Fig.\ref{fig:bilayer}). The two $e_g$ orbitals, $|d_{3z^2-r^2}\rangle,|d_{x^2-y^2}\rangle$ are not split in the trigonal environment.Standard Slater-Koster construction\cite{PhysRev.94.1498} gives the following nearest neighbor(NN) tight-binding model ($t$ is the $dd\sigma$ bond):
\begin{align}
 &H_{TB}=-\sum_{\langle\vec r,\vec r'\rangle,\sigma}\sum_{ab}t^{ab}_{\vec r\;\vec r'}d^{\dagger}_{\vec r,a,\sigma}d_{\vec r',b,\sigma}\notag\\
&t_{\vec r,\vec r\pm \hat x}=\frac{t}{4}\begin{pmatrix}1&-\sqrt{3}\\-\sqrt{3}&3\end{pmatrix},\;\;t_{\vec r,\vec r\pm \hat y}=\frac{t}{4}\begin{pmatrix}1&\sqrt{3}\\\sqrt{3}&3\end{pmatrix},\notag\\
&t_{\vec r,\vec r\pm \hat z}=t\begin{pmatrix}1&0\\0&0\end{pmatrix}.\label{eq:TB}
\end{align}
Here $\vec r,\vec r'$ label the positions of Ni, $a,b$ label the orbital degrees of freedom, and $\sigma$ labels spin. 

\begin{figure}
 \includegraphics[width=0.47\textwidth]{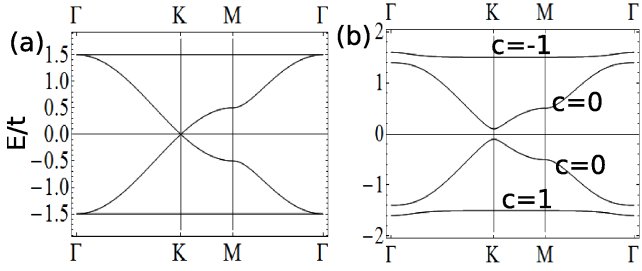}
\caption{(a) The band structure of the NN tight-binding model Eq.(\ref{eq:TB}), each band is two-fold spin-degenerate. The lowest flat band is fully filled. This band structure can also be interpreted as the one of the majority spin in the spin polarized DHSM phase found in our mean-field study, in which case unoccupied minority spin bands are not shown and each majority band is non-degenerate. Fermi level is at the Dirac point. (b) The band structure of the spin polarized QAHI phase at $M_{0,y,0}=0.1t$. Bands' Chern numbers are shown.}
\label{fig:band_structure}
\end{figure}

The four bands (two sublattices and two orbitals) are shown in Fig.\ref{fig:band_structure}(a), including two flat bands on the top and bottom of the spectra, quadratically touching with two dispersive bands in the middle at the $\Gamma$ point. Similar to graphene, the two dispersive bands are linearly touching each other at the $K$ and $K'$ points, the corners of the Brillouin zone. For Ni$^{3+}$ ion with $d^7$ configuration, the $t_{2g}$ is fully filled and the $e_g$ is 1/4 filled, so that the fermi level is positioned at the bottom quadratic band touching point in the non-interacting limit. 

It is well-known that the spin-orbit coupling in the $e_g$ orbitals is zero at the leading order due to the quenched angular momentum. The spin-orbit coupling can be introduced in the $e_g$ orbitals by higher order contributions in a trigonal environment. But because $H_{SO}=\lambda\vec L\cdot\vec S$ is weak for the $3d$ Ni$^{3+}$ ion: $\lambda\sim 80$meV \cite{Vijayakumar199615}, simple estimate from the second order perturbation shows  very small effective spin-orbit coupling in the $e_g$ orbitals $<1$meV. We therefore do not include the atomic spin-orbit coupling in the tight-binding model.

\begin{figure}
\includegraphics[width=0.49\textwidth]{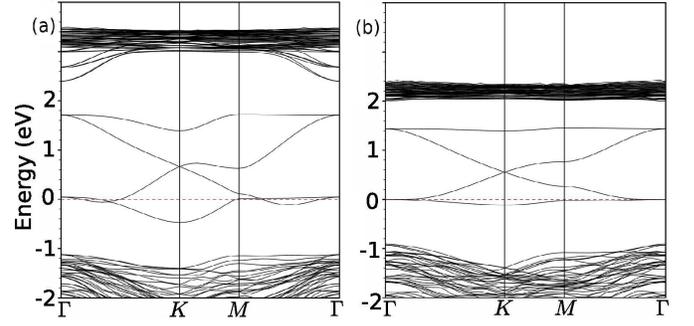}
\caption{Results from the non-magnetic LDA+U calculations of the (a) LaAlO$_3$/LaNiO$_3$/LaAlO$_3$ and (b) LaScO$_3$/LaNiO$_3$/LaScO$_3$ TMOH with lattice relaxation of the LaNiO$_3$ bilayer.}
\label{fig:LDA}  
\end{figure}

Since the $3d$ orbitals are quite confined in space, further neighbor hoppings are suppressed and this NN tight-binding model should be a rather faithful description of the non-interacting electronic structure. 
In Fig.\ref{fig:LDA} we present results from the first-principle GGA+U calculations of the LaAlO3/LaNiO3/LaAlO3 and LaScO3/LaNiO3/LaScO3 bilayer TMOH. The calculations were performed by employing VASP code ~\cite{PhysRevB.54.11169} in the context of density functional theory with the projector augmented wave (PAW) method ~\cite{PhysRevB.50.17953,PhysRevB.59.1758} for the atomic cores and the generalized gradient approximation (GGA) ~\cite{PhysRevLett.77.3865} for exchange-correlation. The GGA+U method was used to treat the 3d electrons of Ni with the Hubbard on-site Coulomb interaction parameter U=7.0 eV and J=0.65 eV ~\cite{0953-8984-9-4-002}. With magnetism suppressed, these band structures are consistent with the NN tight-binding model, with $t\sim 0.6$eV for LaAlO$_3$/LaNiO$_3$/LaAlO$_3$.

The correlation on the Ni$^{3+}$ ions is expected to be intermediate or strong. We consider the standard form of the on-site interactions:
\begin{align}
 &H_{I}=U\sum_{i,a}n_{ia\uparrow}n_{ia\downarrow}+J\sum_{i,a< b}^{\alpha,\beta=\uparrow,\downarrow} d_{ia,\alpha}^{\dagger}d_{ib,\beta}^{\dagger}d_{ia,\beta}d_{ib,\alpha}\notag\\
&+U'\sum_{i,a< b} n_{ia}n_{ib}+J\sum_{i,a<
b}\big(d_{ia,\uparrow}^{\dagger}d_{ia,\downarrow}^{\dagger}d_{ib,\downarrow}d_{ib,\uparrow}+h.c.\big)\notag\\
\label{eq:5-band-model}
\end{align}
$U,U'$ are intra-orbital and inter-orbital Coulomb repulsions and $J$ is the Hund's coupling. In our calculations below, for simplicity, we have set $U'=U-2J$, an equality in rotational symmetric systems. 

The on-site $U\sim 6-7$eV has been used in LDA+U calculations for the nickelates (e.g. \cite{PhysRevB.60.15674}). However Ni oxides have strong charge-transfer effects\cite{PhysRevLett.55.418} to the oxygens. Our model $H=H_{TB}+H_I$ should be treated as an effective model. The values of $U,J$ should be in a range such that the system has an intermediate to strong correlation.

\emph{Symmetries---} Here we only consider the inversion symmetric case with the same substrate on both sides of the sample. The full lattice point group is $D_{3d}$. Apart from the translational symmetry, the symmetry group of the system is $D_{3d}\times SU(2)_{spin}\times TR$ where $TR$ is time-reversal. The band touching points, both the quadratic ones and the linear ones, are protected by this group.

\emph{Mean-field calculation---} We have carried out the variational mean-field study of the model $H=H_{TB}+H_I$. After choosing $J/U=0.2$, a reasonable ratio, the phase diagram as a function of $U/t$ is systematically investigated. To be precise, we introduce the mean-field Hamiltonian:
\begin{align}
 H_{MF}=H_{TB}+\sum_{i,\alpha\beta\gamma}M_{\alpha\beta\gamma}d_{i}^{\dagger}\;\tau_{\alpha}\mu_{\beta}\sigma_{\gamma}\;d_{i},
\end{align}
where $i$ labels unit cells, and real numbers $M_{\alpha\beta\gamma}$, with $\alpha,\beta,\gamma=0,x,y,z$, are the mean-field parameters. $\tau,\mu,\sigma$ are all two-by-two Pauli matrices (the zeroth components are identity matrices) living in the sublattice, orbital, and spin spaces respectively. $d_i$ is the eight-component fermion operator including all these degrees of freedom. The mean-field ground state of $H_{MF}$, $| MF\rangle$, is used to minimize $\langle MF|H|MF\rangle$ numerically to determine the phase diagram.

Because our model only has on-site interactions, only those mean-field parameters involving $\tau_{0,z}$ are considered. We have classified these order parameters according to the symmetry group $D_{3d}\times SU(2)_{spin}\times TR$, and find that each irreducible representation only shows up once. This means that the order parameters do not have to co-exist.

\begin{figure}
 \includegraphics[width=0.47\textwidth]{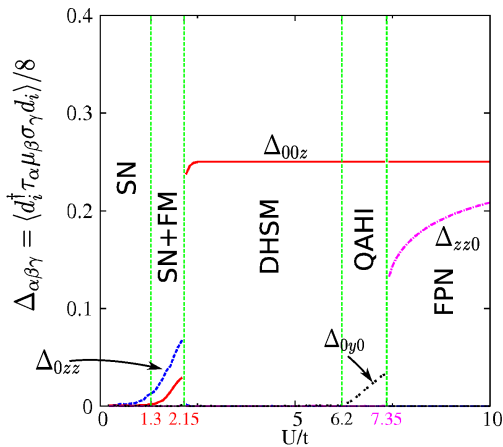}
\caption{(color online) The zero temperature mean-field phase diagram of $H=H_{TB}+H_{I}$ with $U'=U-2J$ and $J/U=0.2$. Vertical axis: the expectation values of the order parameters (defined in the figure, where $d_i$ operator is given in the main text) in these phases. Numerical calculations were performed on a sample with 129$^2$ unit cells.}
\label{fig:phase_diagram}
\end{figure}

The mean-field phase diagram for $J/U=0.2$ at zero temperature is demonstrated in Fig.\ref{fig:phase_diagram}. As $U/t$ is tuned from $0$ to $10$, the phases and phase transitions are summarized in the following. When $U=0$ the chemical potential is at the quadratic band touching point at $\Gamma$ point(see Fig.\ref{fig:band_structure}(a)). A spin nematic phase (SN) occurs at weak $U/t$. After a small region of coexisting with an unsaturated ferromagnetism, a first order transition drives the system into a fully polarized Dirac half-semimetal (DHSM) phase (see Fig.\ref{fig:band_structure}(a)), where four bands of the the majority spin are half-filled and the fermi points are at the $K,K'$ points. The spins remain fully polarized for larger $U/t$. Followed by a second order transition, a  QAHI phase emerges(see Fig.\ref{fig:band_structure}(b) for its band structure). Finally after another first order transition, the system is in a fully polarized nematic phase (FPN).

All spin orders are found to be colinear. For the discussion below, it is helpful to introduce a specific symmetry transformation in such colinear phases. Let us assume the order is along the $S_z$ direction. We define $TR^*$ to be a 180 degree spin-rotation (e.g. around the $S_x$ axis) sending $S_z\rightarrow -S_z$ followed by a $TR$ transformation. A usual colinear order respects $TR^*$ symmetry. In fact, a $TR^*$ symmetry-breaking immediately indicates a dynamically generated spin-orbit coupling.

These phases are characterized by their order parameters and symmetry breaking. The SN phase ($M_{0,z,z}\neq 0$) breaks $D_{3d}$, $SU(2)_{spin}$ and $TR$ symmetries. The DHSM phase ($M_{0,0,z}\neq 0$) breaks both $SU(2)_{spin}$ and $TR$. In the QAHI phase ($M_{0,0,z}\neq 0, M_{0,y,0}\neq 0$)\footnote{$M_{0,y,0}$ and $M_{0,y,z}$ are physically the same here due to the full polarization.}, $SU(2)_{spin}$, $TR$ and even $TR^*$ are broken. QAHI phase is the only phase in the phase diagram breaking $TR^*$ and has dynamically generated spin-orbit coupling. Finally the FPN phase ($M_{0,0,z}\neq 0, M_{z,z,0}\neq 0$)\footnote{$M_{z,z,0}$ and $M_{z,z,z}$ are physically the same here due to the full polarization.} breaks $D_{3d}$, $SU(2)_{spin}$ and $TR$.

The QAHI phase, a band insulator with topologically protected chiral edge metallic modes, is also characterized by a topological index --- the Chern number or TKNN index\cite{PhysRevLett.49.405}.  The total Chern number of this phase is one (see Fig.\ref{fig:band_structure}(b)), which dictates quantized Hall conductance $\sigma_{xy}=\frac{e^2}{h}$ in the ground state. Such a dynamically generated QAHI in an $SU(2)_{spin}$ symmetric Hamiltonian was proposed before\cite{PhysRevLett.100.156401,PhysRevB.79.245331}.

We have found a dominant fully-polarized ferromagnetic order over the majority of the phase diagram. This tendency may be viewed as the residual of the bulk magnetic order ``up-up-down-down'' pattern, and can be qualitatively understood based on the large density of states from the flat band. Therefore we believe that it could be a reliable prediction of this mean-field investigation. The SN phase occurs at weak correlation and is unlikely to be realized. Note that the bandwidth of the bilayer system is substantially smaller than that of the bulk system due to coordination number reduction, and consequently the correlation in the bilayer should be even stronger than that of the bulk. This leads us to believe that \emph{only the DHSM, QAHI and FPN phases are within the reasonable regime of the real material}. If DHSM or QAHI is found experimentally, it will be the first realization of such novel phases of matter.

\emph{Concluding remarks---} We have carried out a systematic mean-field study of the phase diagram of LaNiO$_3$ bilayer TMOH, grown along the (111) direction. We hope that this study could encourage the experimental growths and characterizations of this system. Several interesting candidate quantum phases are found. Among them, the DHSM phase, similar to a spinless graphene, hosts symmetry-protected 2D Dirac cones. This phase has anomalous responses to an orbital magnetic field and can be detected by e.g. quantum oscillation experiments. 

Naively all the spin ordered phases could be destroyed at a finite temperature due to the Mermin-Wagner theorem. But the correlation length of the O(3) nonlinear sigma model, a legitimate model describing the magnetic order fluctuations, diverges exponentially at low temperature. This indicates that even an extremely small atomic spin-orbit coupling, which still exist in the LaNiO$_3$, could pin the order direction and support a rather high temperature phase. 

The tendency of developing a QAH gap in the DHSM phase can be qualitatively understood. At low energy, the only two possible gap terms in the DHSM are CDW ($M_{z,0,0}$) and QAH ($M_{0,y,0}$). CDW is disfavored by inter-orbital repulsion $U'$ so the natural gapped phase continuously connected to the DHSM should be QAHI.

The ferromagnetic order in the QAHI could form smooth textures---the skyrmions since $\pi_2(S^2)=Z$. The dynamical generated spin-orbit coupling, which is an additional $Z_2$ order parameter labeling the breaking of the $TR^*$ symmetry, also could form spatial domain walls. These topological objects could lead to novel physics. We point out that, similar to the quantum Hall ferromagnets\cite{PhysRevB.47.16419}, the skyrmion here is topologically bound with an electric charge and thus is a fermion.

We have not discussed the possible Mott insulator phases with charge fluctuations completely suppressed. This possibility cannot be ignored particularly because the bandwidth of the bilayer is reduced significantly compared with the bulk LaNiO$_3$. Deep in the Mott regime, our model Hamiltonian is reduced to the Kugel-Khomskii-type\cite{KK_model} model whose leading terms favor ferromagnetism on the mean-field level (e.g. see Eq.(2.7) in Ref\cite{PhysRevB.55.8280}). If the full spin polarization persists in this regime, a simple $t/U'$ expansion gives the NN model of orbital fluctuations: $H_{Mott}=J\sum_{\vec r\in A}(\mu^a_{\vec r}\mu^a_{\vec r+\hat x}+\mu^b_{\vec r}\mu^b_{\vec r+\hat y}+\mu^c_{\vec r}\mu^c_{\vec r+\hat z})$, where $A$ labels one sublattice, $J=\frac{t^2}{2U'}>0$ and $\mu^c=\mu_z,\mu^{a,b}=-\frac{1}{2}\mu_z\mp\frac{\sqrt{3}}{2}\mu_x$.  This quantum model of pseudospins $\mu_{\vec r}$, somewhat similar to the Kitaev model\cite{Kitaev20062}, has been used to describe multiferroic layered iron oxides\cite{PhysRevB.78.024416} and has not been solved in a controlled fashion. In a mean-field-type study carried out here, obviously the anti-alignment of the orbital pseudospins is preferred which is exactly the characteristic of the FPN phase. However, quantum fluctuation could lead to exotic phases of matter and this forms a subject of future investigation.

Finally we remark on some other possible phases at weak correlation. In fact, the quadratic band touching point is known to be unstable towards interactions\cite{PhysRevLett.103.046811}. Apart from the nematic phase found here, this instability could lead to a quantum spin Hall insulator (where the spin-orbit coupling also comes from the spontaneous symmetry breaking), or a QAHI with Chern number two. Our mean-field calculation at $J/U=0$ has found these alternative phases at small couplings ($U/t\lesssim 2$). After we finished this manuscript, we have noted a related work\cite{2011arXiv1109.1297R} focusing on these interesting phases.

\emph{Acknowledgement---} Y.R. thanks Andrej Mesaros for helpful discussions. K.Y.Y. and Z.W. are funded by 
DOE-DE-SC0002554. W.Z., D.X., and S.O. are supported by the U.S. Department of Energy, Office of Basic Energy Sciences, Materials Sciences and Engineering Division.

\bibliographystyle{apsrev}
\bibliography{/home/ranying/downloads/reference/simplifiedying}
\end{document}